\documentclass[conference]{IEEEtran}
\newlength{\figurewidth}
\usepackage[pdftex]{graphicx}
\graphicspath{{./figures/}}
\usepackage[usenames,dvipsnames]{xcolor}

\usepackage{amsmath,amssymb,amsthm,amsfonts}
\usepackage{nicefrac}
\usepackage{cite}
\usepackage{color}
\usepackage{transparent}

\ifCLASSOPTIONcompsoc
\usepackage[caption=false,font=normalsize,labelfont=sf,textfont=sf]{subfig}
\else
\usepackage[caption=false,font=footnotesize]{subfig}
\fi

\newcounter{tempEquationCounter}
\newcounter{thisEquationNumber}

\allowdisplaybreaks

\newcommand{\defined}{{\;\overset{\Delta}{=}\;}}

\newcommand{\nd}{n}

\newcommand{\Ktot}{U}
\newcommand{\Kavg}{\overline{K}}
\newcommand{\Nt}{M}
\newcommand{\suf}{T}
\newcommand{\sufss}{^{(\suf)}}

\newcommand{\Ptot}{P_\Sigma}
\newcommand{\Set}{\mathcal{K}}
\newcommand{\wtotal}{w}
\newcommand{\wnormal}{w}

\newcommand{\power}{\rho}

\newcommand{\fixpower}{\bar{\rho}}

\newcommand{\aslot}{\mathsf{A}}
\newcommand{\bslot}{\mathsf{B}}

\newcommand{\expected}[1]{\mathbb{E}\left[#1\right]}

\newcommand{\Prob}[1]{\mathbb{P}\left\{#1\right\}}

\newcommand{\LL}{\mu}

\newcommand{\svec}{\mathbf{s}}

\newcommand{\xvec}{\mathbf{x}}

\newcommand{\yvec}{\mathbf{y}}

\newcommand{\zvec}{\mathbf{z}}
\newcommand{\Hmat}{\mathbf{H}}
\newcommand{\Pmat}{\mathbf{P}}
\newcommand{\Vmat}{\mathbf{V}}
\newcommand{\Ximat}{\mathbf{\Xi}}
\newcommand{\MMPI}{\dag}
\newcommand{\herm}{\mathsf{H}}
\newcommand{\trace}[1]{\mathrm{tr}\left(#1\right)}

\newcommand{\Mellin}{\mathcal{M}}

\newcommand{\s}{{\theta}}
\newcommand{\salt}{\tilde{\theta}}

\newcommand{\ta}{{t_1}}
\newcommand{\tb}{{t_2}}
\newcommand{\ts}{{t}}
\newcommand{\sts}{_t}
\newcommand{\stsb}{_t}
\newcommand{\tsk}{t,}

\newcommand{\Mfun}[1]{\mathbb{K}\left(#1\right)}

\newcommand{\Mfunsuper}[1]{\mathbb{K}^{(T)}\left(#1\right)}

\newcommand{\pv}{{p_\mathrm{v}}}
\newcommand{\pvk}{{p_{\mathrm{v},k}}}

\newcommand{\Abit}{\mathit{A}}

\newcommand{\Sbit}{\mathit{S}}
\newcommand{\Abitcum}{{\mathbf{A}}}
\newcommand{\Dbitcum}{{\mathbf{D}}}
\newcommand{\Sbitcum}{{\mathbf{S}}}

\newcommand{\Delay}{{\mathit{W}}}
\newcommand{\Asnr}{\mathcal{A}}

\newcommand{\Ssnr}{\mathcal{S}}

\newcommand{\Xsnr}{\mathcal{X}}

\IEEEoverridecommandlockouts 
\begin{document}
\title{Delay Performance of the Multiuser MISO Downlink}

\author{
  \IEEEauthorblockN{Sebastian~Schiessl\IEEEauthorrefmark{1}, James~Gross\IEEEauthorrefmark{1} and Giuseppe~Caire\IEEEauthorrefmark{3}}
  \IEEEauthorblockA{\IEEEauthorrefmark{1}School of Electrical Engineering and Computer Science, KTH Royal Institute of Technology, Stockholm, Sweden}
  \IEEEauthorblockA{\IEEEauthorrefmark{3} Institute for Telecommunication Systems, Technical University Berlin, Berlin, Germany}
  Emails: $\{$schiessl,jamesgr$\}$@kth.se, and caire@tu-berlin.de
}
\date{\today}
\maketitle

\newtheorem{lemma}{Lemma}
\newtheorem{assumption}{Assumption}
\newtheorem{corollary}{Corollary}


\begin{abstract}
We analyze a MISO downlink channel where a multi-antenna transmitter communicates with a large number of single-antenna receivers. Using linear beamforming or nonlinear precoding techniques, the transmitter can serve multiple users simultaneously during each transmission slot. However, increasing the number of users, i.e., the multiplexing gain, reduces the beamforming gain, which means that the average of the individual data rates decreases and their variance increases. We use stochastic network calculus to analyze the queueing delay that occurs due to the time-varying data rates. Our results show that the optimal number of users, i.e., the optimal trade-off between multiplexing gain and beamforming gain, depends on incoming data traffic and its delay requirements.
\end{abstract}


\section{Introduction}
\label{sec:introduction}
The capacity of wireless communication systems can be significantly increased when both the transmitter and the receiver are equipped with multiple antennas.
Interestingly, similar capacity gains can also be achieved when a multi-antenna transmitter communicates simultaneously with multiple receivers that have only a single antenna each. 
In order to achieve the capacity of such a multi-user multiple-input single-output (MU-MISO) downlink channel, the transmitter must employ nonlinear precoding techniques like dirty paper coding (DPC) \cite{caire2003achievable}. 
However, linear precoding techniques are sufficient to achieve a large fraction of the capacity. 
A commonly used linear precoding scheme is zero-forcing beamforming (ZFBF), which projects the signal intended for a user into a subspace that is orthogonal to the channels of the other users.
A transmitter with $M$ antennas can use ZFBF to send $K\le M$ different data streams to $K$ users at a time. Increasing $K$ increases the multiplexing gain, but it decreases the beamforming gain due to a reduced dimensionality of the subspaces that are orthogonal to the other users. 
Thus, when $K$ becomes equal to $M$, the linear growth in capacity is lost \cite{peel2005vector}.
Previous works, e.g. \cite{hochwald2002space}, have studied the optimal number of scheduled users $K$, i.e., the optimal trade-off between the multiplexing gain and beamforming gain such that the ergodic capacity of the system is maximized.

However such an analysis of the ergodic sum capacity does not accurately reflect the performance when the system is subject to constraints on maximum delay, such as in live video or audio transmissions. This is due to two reasons. First, when the total number of users $\Ktot$ is larger than the number of antennas $\Nt$, then the transmitter can only schedule a subset of $K<\Ktot$ users in each transmission slot. In order to meet strict requirements on the delay, the scheduling scheme must ensure that each user is scheduled regularly. Second, large variations in the instantaneous data rates mean that the transmitter cannot always send all the available data. When the channel conditions are bad, the data must be stored in a buffer for transmission in subsequent time slots, causing a buffering or queueing delay. 

\subsection{Related Work}
Several works have studied the use of linear precoding in the multiuser MISO downlink, as nonlinear dirty paper coding techniques are difficult to implement in practice. 
Specifically, Yoo and Goldsmith \cite{yoo2006optimality} showed that when the total number of users $\Ktot$ in the system greatly exceeds the number of antennas, then ZFBF achieves asymptotically the same performance as DPC. 
However, their scheme assumes that the transmitter has channel state information (CSI) of all users. The cost of collecting this CSI would be overwhelming when the number of users $\Ktot$ is large.
Sharif and Hassibi \cite{sharif2005capacity} reduce the overhead from collecting CSI by randomly creating a set of beamforming vectors and then transmitting only to the users which report the highest signal-to-interference-and-noise ratio (SINR) along those random beams. Although the scheduling probabilities of all users are equal, the scheduling of the users is random, which can result in unacceptably long delays for some users.
Zhang et al. \cite{zhang2011multi} studied the optimal number of scheduled users when the transmitter has only knowledge of the channel of the scheduled users, and also considered imperfect CSI. Ravindran and Jafar \cite{ravindran2012multiuser6247447} also studied imperfect CSI due to quantized feedback. They found that collecting many bits of feedback (accurate CSI) from very few users is more beneficial than collecting few bits of feedback from many users, which supports the assumption in \cite{zhang2011multi} that CSI should be obtained only for the scheduled users.

However, all of these works studied only the ergodic capacity of the MU-MISO downlink, and did not address the system performance under delay constraints. 
When the transmission rate varies due to channel fading, the transmitter cannot always transmit all data and must keep data in a buffer, causing a random queueing delay.
This queueing delay can be analyzed through the frameworks of stochastic network calculus  \cite{fidler2006network,alzubaidy2016ton} or effective capacity \cite{wu2003effective}. 
Several authors \cite{liu2008effective4595458,jorswieck2010effective5550915,gursoy2011mimo6006584} studied the effective capacity of MIMO systems considering only the single user case. 
Li et al. \cite{li2013adaptive} investigated the effective capacity of multiuser MIMO systems. However, the authors make many assumptions that we do not consider practical, e.g., that the channel coefficients are non-fading and that there is always a backlog of data in each user's queue.

\subsection{Contributions}
In this paper, we analyze the queueing performance of the MU-MISO downlink using stochastic network calculus (SNC). We consider both linear ZFBF precoding and nonlinear dirty paper coding. We demonstrate that SNC can still be applied when the users are not scheduled in each transmission slot, but scheduled regularly in a round robin fashion. Based on previous results, we present closed-form expressions to analytically determine the distribution of the queueing delay. Our numerical results show that the optimal number of scheduled users changes when considering the delay performance instead of the ergodic capacity.

This paper is structured as follows: The system model is given in Sec.~\ref{sec:system_model}. 
In Sec.~\ref{sec:analysis}, we briefly summarize SNC and then derive analytical delay bounds for the considered scenarios. We present numerical evaluations in Sec.~\ref{sec:numerics} and our conclusions in Sec.~\ref{sec:conclusions}.

\section{System Model}
\label{sec:system_model}
We consider downlink transmissions in a time-slotted system from a single base station with $\Nt$ antennas to $\Ktot$ users. We consider the case $\Ktot>\Nt$, where the transmitter cannot serve all users at once. Instead, in each time slot $\ts$, only a subset $\Set_\ts\subset \{1,\ldots,\Ktot\}$ of users are scheduled for transmission, with $K_\ts\defined|\Set_\ts|\leq \Nt$. 
Contrary to \cite{yoo2006optimality}, we assume that the scheduling scheme does not depend on the channel states, as acquiring channel state information (CSI) for all $\Ktot$ users would result in an infeasible amount of overhead. Instead, we follow \cite{zhang2011multi}, where the channel is estimated only for the scheduled users. We assume that the base station has perfect CSI for all $K_\ts$ scheduled users. 

We describe in Sec.~\ref{ssec:system_data_phy} the physical layer transmission for the scheduled users $\Set_\ts$. Round robin scheduling is presented in Sec.~\ref{ssec:link_layer_strategies}. Then, we describe in Sec.~\ref{ssec:system_link} the queueing delay of the system on the link layer, followed by the problem statement in Sec.~\ref{ssec:problem_statement}.

\subsection{Physical Layer Model}
\label{ssec:system_data_phy}
The received signal $\yvec\sts\in\mathbb{C}^{K_\ts\times 1}$ at the $K_\ts$ scheduled users in time slot $\ts$ can be described as
\begin{equation}
\yvec\sts = \Hmat\sts \xvec\sts +\zvec\sts
\,.
\end{equation}
For the channel matrix $\Hmat\sts\in\mathbb{C}^{K_\ts \times \Nt}$, we assume Rayleigh fading, i.e., all elements are independent and identically distributed (i.i.d.) with Gaussian distribution $\mathcal{CN}(0,1)$. Furthermore, we consider the quasi-static fading model where the channel $\Hmat_\ts$ remains constant for the duration of time slot $\ts$, consisting of $\nd$ channel uses, and changes to an independent realization in the next time slot (note that the set $\Set_\ts$ of scheduled users also changes). The input signal is denoted as $\xvec\sts\in\mathbb{C}^{\Nt \times 1}$ and must satisfy a short-term power constraint $\trace{\expected{\xvec\sts\xvec\sts^\herm}}\le \Ptot$ for each realization of $\Hmat_\ts$. The noise $\zvec\sts \in\mathbb{C}^{K_\ts\times 1}$ has i.i.d. components $\mathcal{CN}(0,1)$. 

Given the channel matrix $\Hmat_\ts$, the transmitter must encode the data for the $K_\ts$ scheduled users into coded symbols $\xvec_\ts$. We now present two different encoding strategies.

\subsubsection{Zero-Forcing Beamforming (ZFBF)}
When the transmitter applies ZFBF, the input signal vector $\xvec\stsb$ is given by \cite{caire2003achievable}
\begin{equation}
	\xvec\stsb =\Vmat\stsb\Pmat\stsb^{1/2}\svec\stsb
\end{equation}
where $\Vmat\stsb$ is the precoding matrix, $\Pmat\stsb=\mathrm{diag}(\rho_{\tsk 1},\ldots,\rho_{\tsk K})$ is the power allocation matrix, and $\svec\stsb$ is the $K\stsb\times 1$ vector of (independently) coded Gaussian symbols for the $K\stsb$ scheduled users.
The ZFBF precoder is given as \cite{caire2003achievable}
\begin{equation}
	\Vmat\stsb=\Hmat\stsb^\herm(\Hmat\stsb\Hmat\stsb^\herm)^{-1} \Ximat\stsb^{1/2}
\end{equation}
where $\Hmat\stsb^\MMPI = \Hmat\stsb^\herm(\Hmat\stsb\Hmat\stsb^\herm)^{-1}$ is the Moore-Penrose pseudo-inverse of $\Hmat\stsb$ and $\Ximat\stsb=\mathrm{diag}\left(\xi_{\tsk 1},\ldots,\xi_{\tsk K}\right)$ is the normalization matrix such that the columns of $\Vmat\stsb$ have unit-2 norm. The variables $\xi_{\tsk k}$ are central chi-square distributed (scaled by a factor $1/2$) with $2 m\stsb$ degrees of freedom, where $m\stsb = \Nt-K\stsb+1$. Their PDF is given by \cite[Lemma~4]{caire2003achievable}
\begin{equation}
	f_m(\xi)=\frac{1}{\Gamma(m)} \xi^{m-1} e^{-\xi}
	\label{eq:pdf_central_chisquare_scaled}
	\;.
\end{equation}

We asume that the blocklength $\nd$ of the channel code is sufficiently long, so that the system can achieve error-free transmission to user $k$ at a rate \cite{caire2003achievable}
\begin{equation}
	R_{k}(\ts)=
	\log_2(1+\power_{\tsk k}\xi_{\tsk k})
	\,.
	\label{eq:rates_zfbf}
\end{equation}

\subsubsection{Zero-Forcing with Dirty-Paper Coding (ZF-DPC)}
For comparison, we also present a scheme known originally as \emph{ranked known interference (RKI)} \cite{caire2003achievable}.
Assume that the scheduled users $\Set\stsb\subset \{1,\ldots,\Ktot\}$ are ordered from $1$ to $K\stsb$.
When a scheduled user $k\in\Set\stsb$ is the $\kappa$-th ordered user, it experiences interference from the ordered users $1,\ldots, \kappa-1$. The interference from those users is non-causally known at the transmitter. Therefore, the transmitter can employ dirty paper coding (DPC) when encoding the data for the $\kappa$-th ordered user, which allows sending data at the same rate as if no interference was present. 
Furthermore, if the ordered users $\kappa+1,\ldots,K\stsb$ apply zero-forcing (ZF) towards the users $1,\ldots, \kappa$, then they will not interfere with the $\kappa$-th user. Thus, when user $k$ is the $\kappa$-th ordered user, it can achieve a rate $R_{k}(\ts)=
	\log_2(1+\power_{\tsk k}\xi_{\tsk k})$,
where the variables $\xi_{\tsk k}$ have central chi-square distribution (scaled by $1/2$) with $2m_{\tsk k}$ degrees of freedom with $m_{\tsk k}=\Nt-\kappa+1$. The PDF of $\xi_{\tsk k}$ is given by \eqref{eq:pdf_central_chisquare_scaled}.
Note that $m_{\tsk k}$ and the rates $R_{k}(\ts)$ depend on the user ordering.

\subsection{Round Robin (RR) Scheduling}
\label{ssec:link_layer_strategies}
In the considered scenario, the number of users $\Ktot$ exceeds the number of transmit antennas $\Nt$. Therefore, the transmitter must schedule a subset $\Set_\ts$ of users in each time slot $\ts$. We consider round robin (RR) scheduling as in \cite{zhang2011multi}, where multiple users can be scheduled in each time slot.
Each user $k$ is scheduled exactly once within a superframe of $\suf$ slots.
The average number of scheduled users per slot is then given as $\Kavg\defined\Ktot/ \suf$, with $1\le \Kavg \le \Nt$.
As the total number of users $\Ktot$ is fixed, $\Kavg$ may not always be an integer, and thus the scheduler must sometimes select more than $\Kavg$ users, sometimes less.
We assume that in $\suf_\aslot$ of the subslots, $K_\aslot=\left\lceil\, \Kavg\, \right\rceil$ users are served, in $\suf_\bslot=\suf-\suf_\aslot$ of the subslots, $K_\bslot=\lfloor \Kavg \rfloor$ users are served, such that the total number of users served in the superframe is $\suf_\aslot K_\aslot+\suf_\bslot K_\bslot=\Ktot$. 

In order to maintain fairness between the users, the transmitter randomly assigns the users to the slots in each superframe. Furthermore, in case of ZF-DPC, where the performance depends on the encoding order of the users, we require that the users are ordered randomly.

\subsection{Link Layer Model}
\label{ssec:system_link}
In time slot $\ts$, $A_k(\ts)$ data bits intended for downlink transmission to user $k$ arrive at the transmitter. The data is stored in a transmit buffer, with individual buffers (or queues) for each user.
We assume that the arrival process $A_k(\ts)$ is constant over time and equal for all users, with $\alpha$ denoting the constant number of bits that arrive in the queue of each user in each time slot.
The service rate offered by the wireless system in each time slot to a scheduled user $k\in\Set_\ts$ is given by $S_k(\ts)=\nd R_{k}(\ts)$, and $S_k(\ts)=0$ when $k\notin\Set_\ts$. 
The departure process $D_k(\ts)$ describes the amount of data that is transmitted to the receiver. Thus, $D_k(\ts)$ is limited both by the amount of data waiting in the buffer, as well as by the service rate $S_k(\ts)$.
The cumulative arrival, service, and departure processes are defined as
\begin{equation}
\Abitcum_k(\ta,\tb) \defined \sum\limits_{\ts=\ta}^{\tb-1}A_k(\ts) \;,
\quad\Sbitcum_k(\ta,\tb) \defined \sum\limits_{\ts=\ta}^{\tb-1}S_k(\ts) \;,
\end{equation}
\begin{equation}
\quad\Dbitcum_k(\ta,\tb) \defined \sum\limits_{\ts=\ta}^{\tb-1}D_k(\ts) \;.
\end{equation}
The queueing delay $\Delay_k(\ts)$ of user $k$ at time $\ts$ is defined as the time it takes for all data that arrived prior to time $\ts$ to depart from the transmit buffer and reach the receiver \cite{alzubaidy2016ton, schiessl2016imperfectcsi}:
\begin{equation}
\Delay_k(\ts) \defined \inf\left\{u\geq 0:\quad \Abitcum_k(0,\ts) \leq \Dbitcum_k(0,\ts+u) \right\} \;.
\label{eq:def_delay}
\end{equation}
The delay $\Delay_k(\ts)$ is random. We want to find the probability $\pvk(w)$ that the delay $\Delay_k(\ts)$ of the data for user $k$ exceeds a specified target delay $\wtotal$ at any time $\ts$:
\begin{equation}
\pvk(\wtotal) \defined \sup_{\ts\ge 0} \left\{ \Prob{\Delay_k(\ts)>\wtotal} \right\}\; .
\label{eq:def_pdelayviol_time}
\end{equation}

\subsection{Problem Statement}
\label{ssec:problem_statement}
In this work, we want to find the value $\Kavg$ that minimizes the delay violation probability $\pvk(w)$.
On the one hand, choosing a small value of $\Kavg$ means that only few users are scheduled in each time slot, so that their signals are transmitted with high beamforming gain and transmit power. However, this also results in a small multiplexing gain and a long time to schedule all users. On the other hand, a large value $\Kavg$ results in poor beamforming gain. 

We note that the delay violation probability $\pvk(\wtotal)$ cannot be determined directly in an analytically tractable form.
However, the delay violation probability can be analytically approximated/bounded using the frameworks of effective capacity \cite{wu2003effective} or stochastic network calculus \cite{fidler2006network,alzubaidy2016ton}. Effective capacity provides an approximation for $\pvk(\wtotal)$ that is tight for large $\wtotal$. In this work, we perform the optimization of $\Kavg$ based on stochastic network calculus, as it provides a strict upper bound on $\pvk(\wtotal)$ that holds also for small $\wtotal$.

\section{Analysis}
\label{sec:analysis}
In Sec.~\ref{ssec:queueing}, we present a summary of the delay analysis through stochastic network calculus in a transform domain \cite{alzubaidy2016ton}. 
We demonstrate in Sec.~\ref{ssec:queueing_rr} how stochastic network calculus can be used when round robin scheduling is used. In Sec.~\ref{ssec:delay_analysis}, we analytically obtain the stochastic network calculus bounds for the considered scenario. Note that the transmission and scheduling strategies in Sec.~\ref{sec:system_model} are fair, as the distribution of the service process $\Sbit_k(\ts)$ is the same for all users. We assume that all users are subject to the same delay requirements and thus drop the subscript $k$ to shorten the notation.

\subsection{Stochastic Network Calculus (SNC)}
\label{ssec:queueing}
This section closely follows our previous work \cite{schiessl2016imperfectcsi} and provides a summary of stochastic network calculus \cite{fidler2006network,alzubaidy2016ton}.
 
The delay $\Delay(t)$ in (\ref{eq:def_delay}) is defined in terms of the arrival and departure processes. 
However, the distribution of the delay can be found directly from the statistics of the arrival and service processes.
We follow \cite{alzubaidy2016ton} and describe these processes in the exponential domain, also referred to as \emph{SNR domain}. The arrival and service processes in the bit domain, $\Abit(\ts)$ and $\Sbit(\ts)$, are converted to the SNR domain (denoted by calligraphic letters) as 
\begin{equation}
\Asnr(\ts) \defined e^{\Abit(\ts)}\,,\quad\Ssnr(\ts) \defined e^{\Sbit(\ts)}
\,.
\end{equation}
In this work, we assume constant arrivals with $\Abit(\ts)=\alpha$.
Consider for now a service process $\Sbit(\ts)$ that is independent and identically distributed (i.i.d.) between time slots. 
An upper bound on the delay violation probability $\pv(w)$ can then be obtained in terms of the Mellin transforms of $\Asnr$ and $\Ssnr$.
The Mellin transform $\Mellin_\Xsnr(\s)$ of a nonnegative random variable $\Xsnr$ is defined as \cite{alzubaidy2016ton}
\begin{equation}
\Mellin_\Xsnr(\s)\defined\mathbb{E}\left[\Xsnr^{\s-1}\right]
\end{equation}
for a parameter $\s \in \mathbb{R}$. 
For the analysis, we choose $\s>0$ and check if the stability condition $\Mellin_{\Asnr}(1+\s)\Mellin_{\Ssnr}(1-\s) < 1$ holds.
If it holds, define the kernel \cite{alzubaidy2016ton,schiessl2015delay}
\begin{align}
\Mfun{\s,\wnormal} &\defined \lim\limits_{t\to\infty} \sum_{u=0}^{t} \Mellin_{\Asnr}(1+\s)^{t-u} \cdot \Mellin_{\Ssnr}(1-\s)^{t+\wnormal-u}
\nonumber
\\
&= \frac{\Mellin_{\Ssnr}(1-\s)^{\wnormal}}{1-\Mellin_{\Asnr}(1+\s)\Mellin_{\Ssnr}(1-\s)} 
\;.
\label{eq:snc_kernel}
\end{align}
For any parameter $\s>0$, the kernel $\Mfun{\s,\wnormal}$ provides an upper bound on the delay violation probability $\pv(w)$ \cite{alzubaidy2016ton, schiessl2015delay}. This holds for any time slot $t$, including the limit $t\to\infty$ (steady-state). 
In order to find the tightest upper bound, one must find the parameter $\s>0$ that minimizes $\Mfun{\s,\wnormal}$:
\begin{equation}
\pv(\wnormal) \leq \inf_{\s>0}\left\{ \Mfun{\s,\wnormal} \right\} \;.
\label{eq:pdelay_bound}
\end{equation}

\subsection{SNC and Round Robin Scheduling}
\label{ssec:queueing_rr}
For round robin scheduling, the delay analysis through stochastic network calculus as shown in Sec.~\ref{ssec:queueing} cannot be applied directly, as $\Sbit(\ts)$ is zero in the time slots where the user is not scheduled, i.e., $\Sbit(\ts)$ is not i.i.d. between time slots. However, stochastic network calculus can be applied on the superframe level. The service that a user receives in superframe $i$ is denoted as $\Sbit^{(\suf)}(i)$, and is i.i.d. between superframes, because each user is scheduled exactly once per superframe of length $\suf$. The arrival process on the superframe level is given as $\Abit^{(\suf)}(i)= \alpha T$ bits, and the Mellin transform of the process $\Asnr$ in the SNR domain is $\Mellin_{\Asnr^{(T)}}(\s) = e^{\alpha T(\s-1)}$.

Assume first that $\wtotal/\suf$, where $\wtotal$ is maximum delay in time slots, is an integer:  Then, the queueing analysis can easily be done on the superframe level:
\begin{align}
\pv(\wtotal) & \le \Mfunsuper{\s,\frac{\wtotal}{\suf}} 
\label{eq:snc_pv_bound_singlegroup}
\,,
\end{align}
with
\begin{align}
\Mfunsuper{\s,\frac{\wtotal}{\suf}}&= \frac{\Mellin_{\Ssnr^{(\suf)}}(1-\s)^{\frac{\wtotal}{\suf}}}{1-\Mellin_{\Asnr^{(\suf)}}(1+\s)\Mellin_{\Ssnr^{(\suf)}}(1-\s)} 
\,.
\label{eq:snc_kernel2}
\end{align}
In case $\wtotal/\suf$ is not an integer, some users (denoted as group 1) will be served $\lceil \wtotal/\suf\rceil$ times before the deadline, while others (group 2) will only be served $\lfloor\wtotal/\suf\rfloor$ times. For the sake of fairness, we assume that the users are assigned randomly to the slots. Then, the probability of being in the second group is $p_2=\frac{\mod(\wtotal,\suf)}{\suf}$, and $p_1 = 1-p_2$. Thus, the overall bound on the delay violation probability is given by 
\begin{align}
\pv(\wtotal) \le p_1 \Mfunsuper{\s,\left\lceil\frac{\wtotal}{\suf}\right\rceil} + p_2 \Mfunsuper{\s,\left\lfloor \frac{\wtotal}{\suf}\right\rfloor}
\;.
\label{eq:snc_pv_bound_multigroups}
\end{align}
Similar to \eqref{eq:pdelay_bound}, this bound holds for any $\s>0$, such that finding the tightest possible bound requires taking the infimum over \eqref{eq:snc_pv_bound_multigroups} with respect to $\s$.

\subsection{Delay Analysis for MU-MISO Downlink}
\label{ssec:delay_analysis}
The kernel \eqref{eq:snc_kernel2} depends on the Mellin transform of the service $\Ssnr^{(\suf)}$ offered to each user in each superframe. Users are scheduled exactly once in a superframe, so that $\Ssnr^{(\suf)}$ has the same distribution as the service $\Ssnr$ experienced by a scheduled user. The SNR-domain service process of a scheduled user is given as $\Ssnr=e^\Sbit=e^{\nd R}$, with $R=\log_2(1+\power\xi)$.

For ZFBF and ZF-DPC, $\xi$ is a scaled central $\chi^2$ variable with varying degrees of freedom $2m$ as outlined in Sec.~\ref{ssec:system_data_phy}. For ZFBF, we have $m=\Nt-K_{(\aslot/\bslot)}+1$, depending on the slot type $(\aslot/\bslot)$. For ZF-DPC, $m\in\left\{1,\ldots,\Nt-K_{(\aslot/\bslot)}+1\right\}$, each with probability $p_{m|(\aslot/\bslot)} = 1/K_{(\aslot/\bslot)}$.

The transmitter is subject to a short-term power constraint $\trace{\expected{\xvec\stsb\xvec\stsb^\herm}}\le \Ptot$.
A simple power allocation strategy shares $\Ptot$ equally among the $K_\aslot$ or $K_\bslot$ scheduled users:
\begin{align}
\power=\left\{\begin{array}{ll}
\fixpower_\aslot=\frac{\Ptot}{K_\aslot} &\quad \text{with prob.} \quad p_{\aslot}=\frac{K_{\aslot} \suf_{\aslot}}{\Ktot} \\
\fixpower_\bslot=\frac{\Ptot}{K_\bslot} &\quad \text{with prob.} \quad p_{\bslot}=\frac{K_{\bslot} \suf_{\bslot}}{\Ktot}
\end{array}\right.
\label{eq:power_alloc_simple}
\;.
\end{align}

The Mellin transform of the service process $\Ssnr\sufss$ can be obtained by averaging over the Mellin transforms of the service process with specific values of $\fixpower$ and $m$:
\begin{equation}
\Mellin_{\Ssnr\sufss}(1-\s) = \sum_{\fixpower,m}p_{\fixpower,m}\Mellin_{\Ssnr\sufss|\fixpower, m}(1-\s)
\label{eq:mellin_service_fixedpowers}
\,,
\end{equation}
where $p_{\fixpower,m}$ denotes the joint probability of a user's channel having $2m$ degrees of freedom ($\xi\sim\frac{1}{2}\chi^2_{2m}$) and power $\fixpower$.\footnote{For ZFBF, $p_{\fixpower,m}$ is equal to $p_{\aslot}$ or $p_{\bslot}$ as given in \eqref{eq:power_alloc_simple}. For ZF-DPC, the different $p_{\fixpower,m}$ can simply be obtained as $p_{\aslot/\bslot}\cdot p_{m|(\aslot/\bslot)}$.}

For a specific constant power $\fixpower$ and a specific $m$, the Mellin transform of the service process can be obtained as
\begin{align}
\Mellin_{\Ssnr\sufss|\fixpower, m}(1-\s) &=\expected{\left.\left(e^{\nd R}\right)^{-\s}\right| \fixpower, m}
\\&= \expected{\left.(1+\fixpower\xi)^{-\frac{\s \nd}{\ln 2}}\right| m}
\,.
\end{align}
We define $\salt\defined\frac{\s \nd}{\ln 2}$ and follow the derivations in \cite{schiessl2017wsa} to obtain
\begin{align}
&\expected{\left.(1+\fixpower\xi)^{-\salt}\right| m} =\int\limits_0^\infty (1+\fixpower\xi)^{-\salt} f_m(\xi)d\xi 
\\&=\int\limits_0^\infty (1+\fixpower\xi)^{-\salt} \frac{1}{\Gamma(m)} \xi^{m-1} e^{-\xi}d\xi 
\\&=\sum\limits_{\LL=0}^{m-1}\frac{\binom{m-1}{\LL} (-1)^{\LL}}{\Gamma(m)\fixpower^{m-1}} \int\limits_0^\infty (1+\fixpower\xi)^{m-1-\LL-\salt} e^{-\xi}d\xi 
\label{eq:mellin_service_step2}
\\&=\sum\limits_{\LL=0}^{m-1}\frac{\binom{m-1}{\LL} (-1)^{\LL}}{\Gamma(m)} \fixpower^{-\LL-\salt} e^{\frac{1}{\fixpower}}
\nonumber
\\& \quad\cdot \int\limits_0^\infty \left(\frac{1}{\fixpower}+\xi\right)^{m-1-\LL-\salt} e^{-\left(\frac{1}{\fixpower}+\xi\right)}d\xi 
\\&=\sum\limits_{\LL=0}^{m-1}\frac{\binom{m-1}{\LL} (-1)^{\LL}}{\Gamma(m)} \fixpower^{-\LL-\salt} e^{\frac{1}{\fixpower}}
\Gamma\left(m-\mu-\salt,\frac{1}{\fixpower}\right)
\,.
\label{eq:mellin_service_constpower_series}
\end{align}
In \eqref{eq:mellin_service_step2}, we used the conversion \cite{schiessl2017wsa}
\begin{align}
x^{m-1} &= 
\sum\limits_{\LL=0}^{m-1}\binom{m-1}{\LL}\left(1+x\right)^{m-1-\LL}(-1)^{\LL}
\end{align} 
and in \eqref{eq:mellin_service_constpower_series}, we applied the upper incomplete Gamma function
\begin{align}
\Gamma(s,x) = \int_{x}^{\infty}t^{s-1}e^{-t}dt
\,.
\label{eq:incomplete_gamma}
\end{align}
Thus, given the arrival rate $\alpha$ in bits per time slot and a specific choice of superframe length $\suf$ (which determines the average number of scheduled users $\Kavg=\Ktot/\suf$), the upper bound \eqref{eq:snc_pv_bound_multigroups} on $\pv(w)$ can be obtained analytically through \eqref{eq:mellin_service_fixedpowers} and \eqref{eq:mellin_service_constpower_series}.


\section{Numerical Results}
\label{sec:numerics}
\begin{figure}[t!]
	\vspace{2mm}
	\subfloat[\label{subfig:m8_rate}]{%
		\includegraphics[width=0.98\figurewidth]{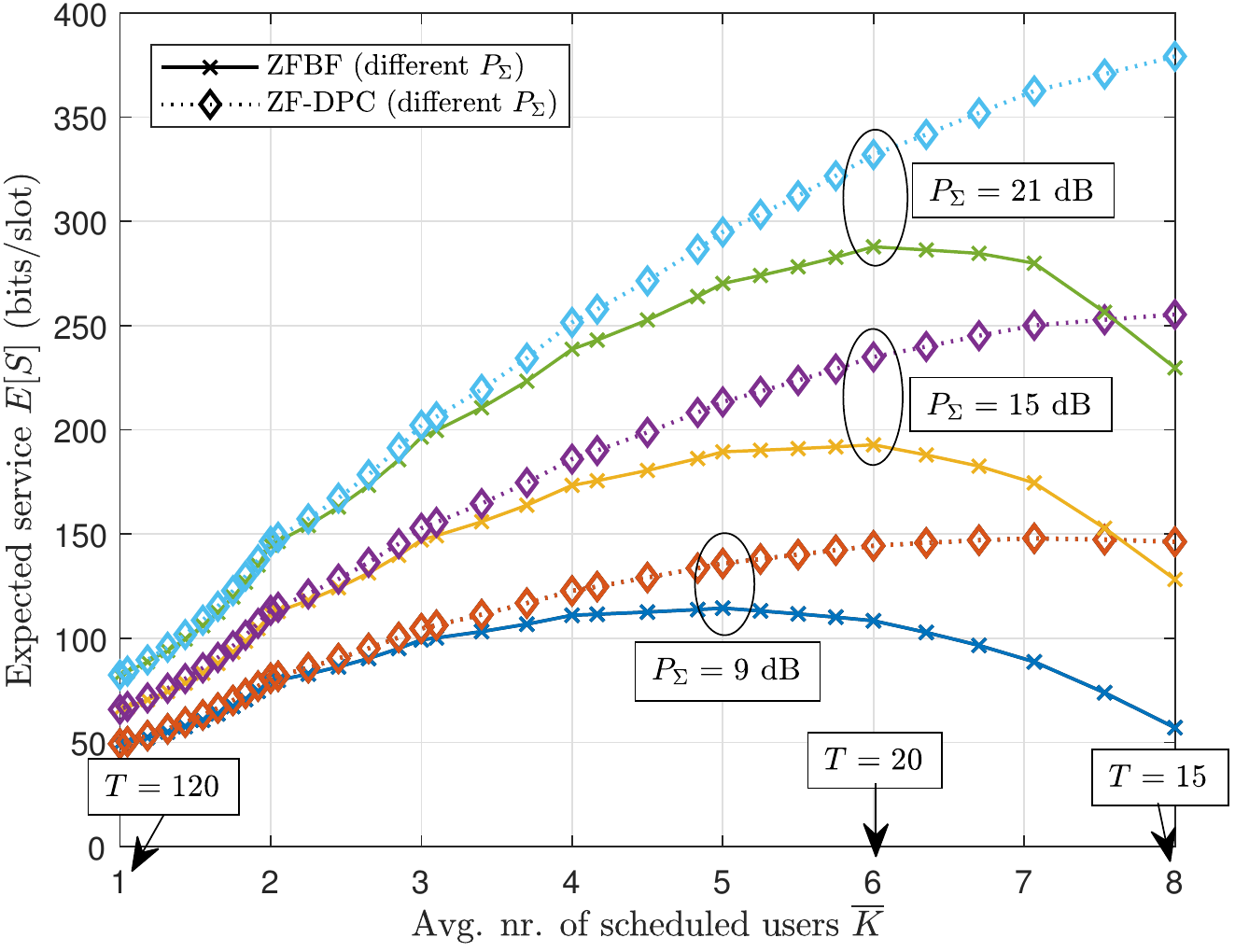}
	}
	\vfill
	\vspace{-0.5mm}
	\subfloat[\label{subfig:m8_zfbf}]{%
		\includegraphics[width=0.98\figurewidth]{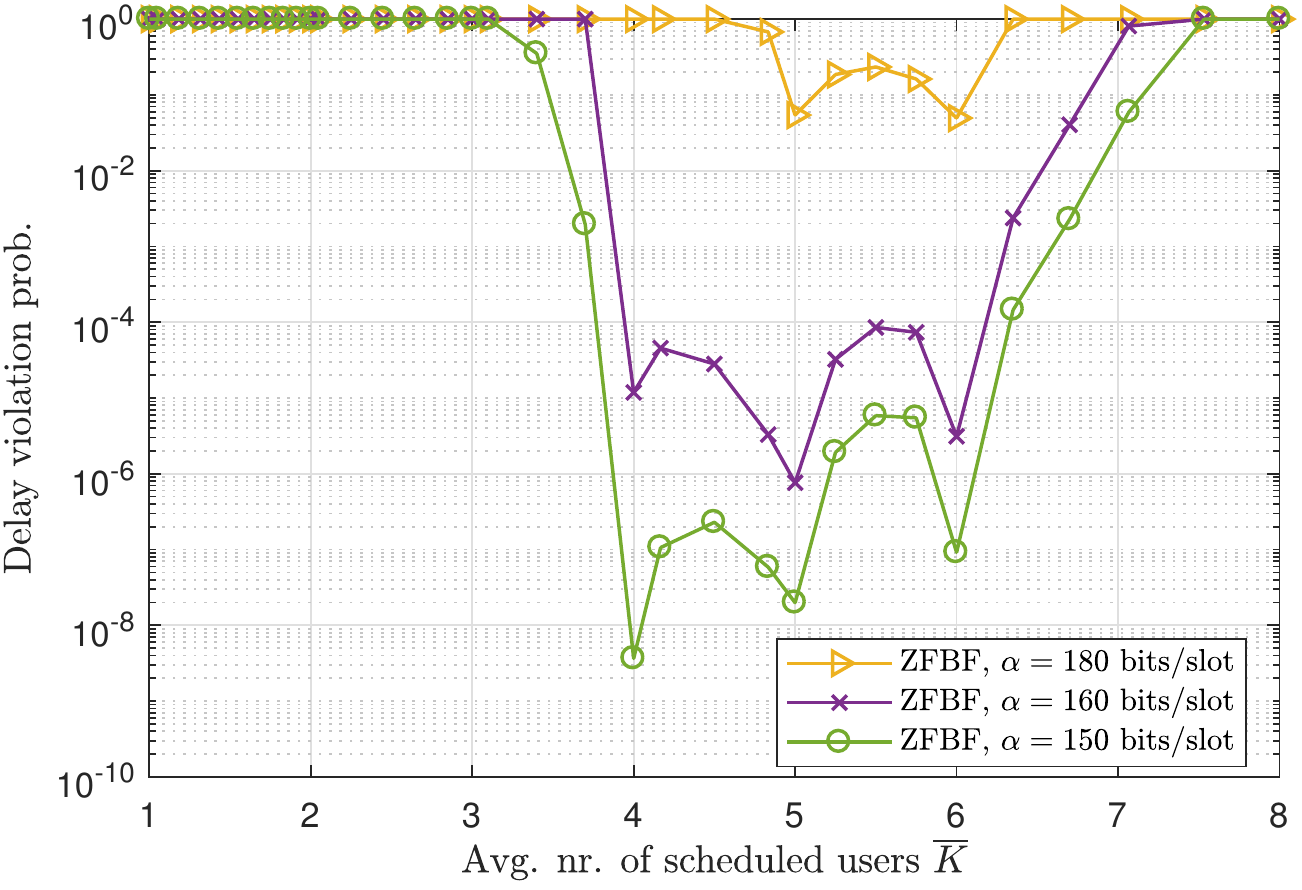}
	}
	\vfill
	\vspace{-0.5mm}
	\subfloat[\label{subfig:m8_dpc}]{%
		\includegraphics[width=0.98\figurewidth]{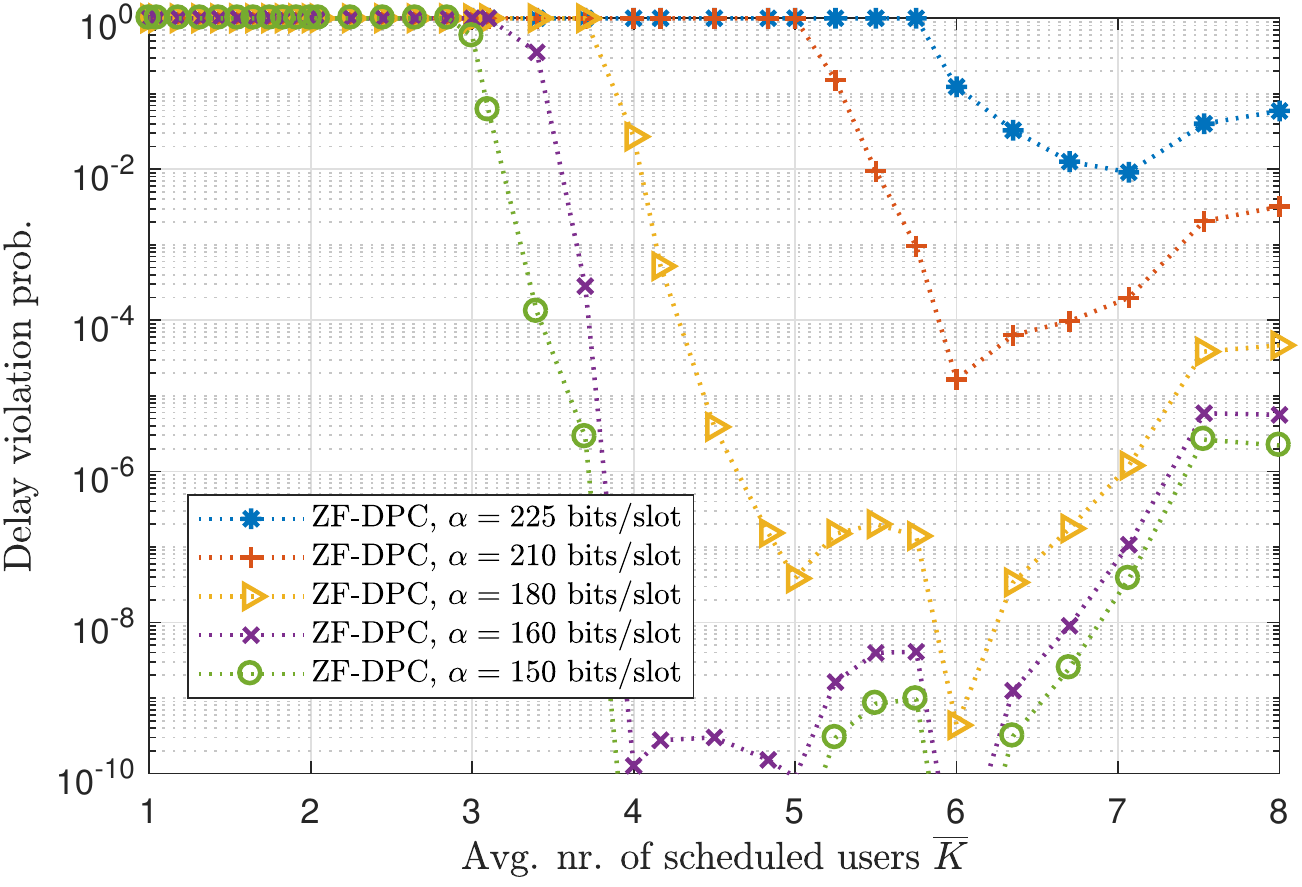}
	}
	\caption{$\Nt=8$, $\Ktot=120$ users, $\nd=1000$ symbols. (a): Expected service rate for  $\Ptot\in\{9,15,21\}~\mathrm{dB}$. (b) Delay violation probability when using ZFBF, according to the SNC bound, for deadline $\wtotal=60$ slots and different arrival rates $\alpha$, for $\Ptot=15~\mathrm{dB}$. (c) same parameters, but using ZF-DPC.}
	\label{fig:results_m8}
\end{figure}

In Fig.~\ref{fig:results_m8}, we show various aspects of the performance of a system with $\Ktot=120$ users and $\Nt=8$ antennas. First, in Fig.~\ref{subfig:m8_rate}, we show the expected service rate per slot vs. the average number of scheduled users $\Kavg$ for different values of the SNR $\Ptot\in\{9,15,21\}~\mathrm{dB}$. Note that the superframe length $\suf$ must always be integer, but $\Kavg=\Ktot/\suf$ is not always integer. In each superframe of $\suf$ time slots, the transmitter sends $nR$ bits to each user. Thus, the expected service rate per user and per time slot is given as
\begin{equation}
\expected{S} = \frac{1}{\suf}\expected{S^{(\suf)}} =\frac{1}{\suf}\expected{nR}
\,.
\end{equation}
For ZFBF, we observe for every SNR $\Ptot$ that the expected service rate first increases and then decreases in $\Kavg$. At very small $\Kavg$, an increase in $\Kavg$ means that more users are scheduled simultaneously, and the multiplexing gain from transmitting to multiple users outweighs the performance loss due to slightly decreased service rates of each user. However, at very large $\Kavg$, the expected service rate decreases, because the relative increase in the number of scheduled users is small, whereas the beamforming gain is massively reduced. 
Furthermore, we observe that the value of $\Kavg$ that maximizes the expected service rate grows with the SNR. This is in line with previous results \cite{hochwald2002space}.
For ZF-DPC, the behavior is different. In fact, the expected service rate is strictly increasing in $\Kavg$ for $\Ptot\in\{15,21\}~\mathrm{dB}$. When adding more users to a ZF-DPC transmission, the additional users do not create any interference towards the previous users.
The only downside from adding more users to the ZF-DPC system is that a small fraction of the transmitted signal power is shared with the new users. 
For very low SNR $\Ptot$, this effect may decrease the expected service at large $\Kavg$, but even at $\Ptot=9~\mathrm{dB}$, this effect remains almost unnoticeable.

In Fig.~\ref{subfig:m8_zfbf} and Fig.~\ref{subfig:m8_dpc}, we consider the delay performance of the system for ZFBF and ZF-DPC, respectively, with different arrival rates $\alpha$, a maximum delay of $w=60$ time slots, and with $\Ptot=15~\mathrm{dB}$. 
For ZFBF, Fig.~\ref{subfig:m8_zfbf} shows that the delay violation probability, obtained from the analytical bound \eqref{eq:snc_pv_bound_multigroups}, remains high at $\alpha=180$ bits/slot. However, when the arrival rate is decreased, the delay violation probability decreases significantly. Interestingly, the minimum in the delay violation probability is attained at $\Kavg=6$ for $\alpha=180$, at $\Kavg=5$ for $\alpha=160$, and at $\Kavg=4$ for $\alpha=150$ bits/slot. Thus, the optimal value of $\Kavg$ changes depending on the arrival rate and delay constraints imposed on the system. Many of our additional experiments also show that the optimal $\Kavg$ under delay constraints is slightly below the value of $\Kavg$ that maximizes the expected service rate. 
An explanation for this phenomenon is that even though decreasing the number of users $\Kavg$ means that users are scheduled less often (lower multiplexing gain), the system has a higher beamforming gain, i.e., the channel gains $\xi$ of all users have more degrees of freedom. This decreases the variance of the service $\Sbit$ experienced by each user and thus improves the delay performance of the system.

Fig.~\ref{subfig:m8_dpc} shows the delay violation probability for ZF-DPC. Here, we observe that the minimum in the delay violation probability is attained at $\Kavg\approx 7$ for $\alpha=225$ and at $\Kavg= 6$ for $\alpha=210$, whereas Fig.~\ref{subfig:m8_rate} showed that the expected service rate is maximized at $\Kavg=8$. The explanation is similar to the explanation in case of ZFBF: When scheduling $\Kavg=8$ users, the effective channel gains $\xi$ for some of the users (the users which are encoded last in the ZF-DPC order) have only 2 degrees of freedom. These users may experience very low data rates, so that the delay violation probability increases.

\begin{figure}[t!]
	\vspace{2mm}
	\hspace{1mm}
	\subfloat[\label{subfig:results_M8_opt_Kavg}]{%
		\includegraphics[width=0.97\figurewidth]{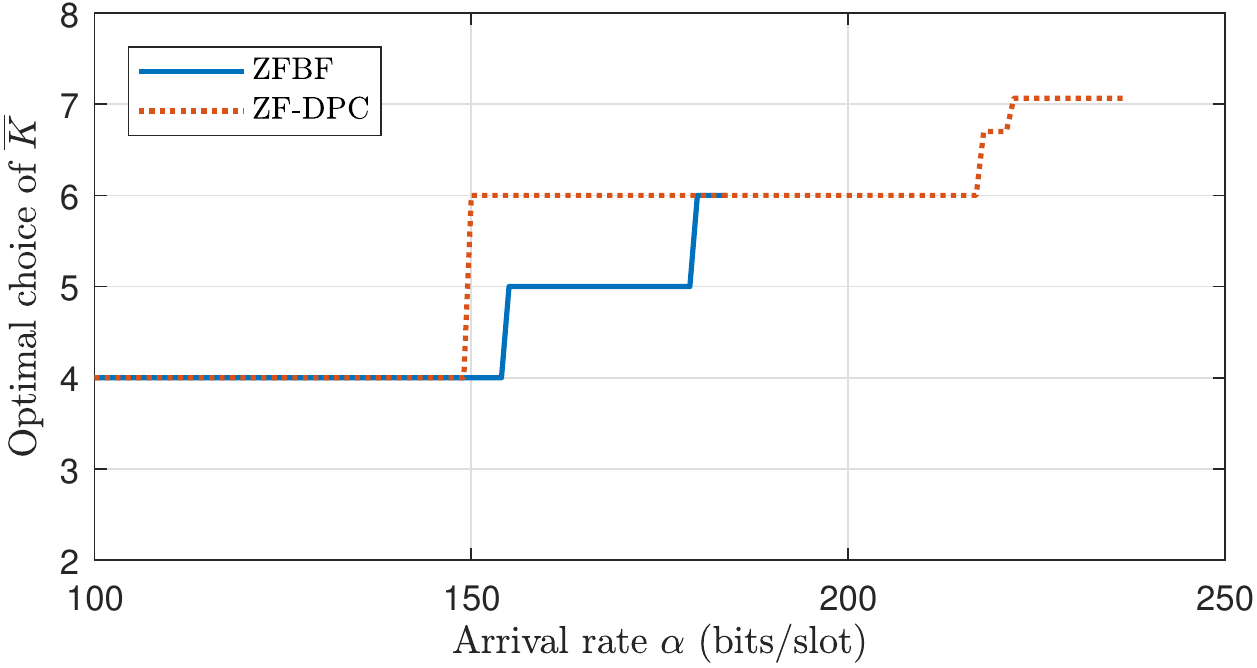}
	}
	\vfill
	\vspace{-2mm}
	\subfloat[\label{subfig:results_M8_min_pv}]{%
		\includegraphics[width=\figurewidth]{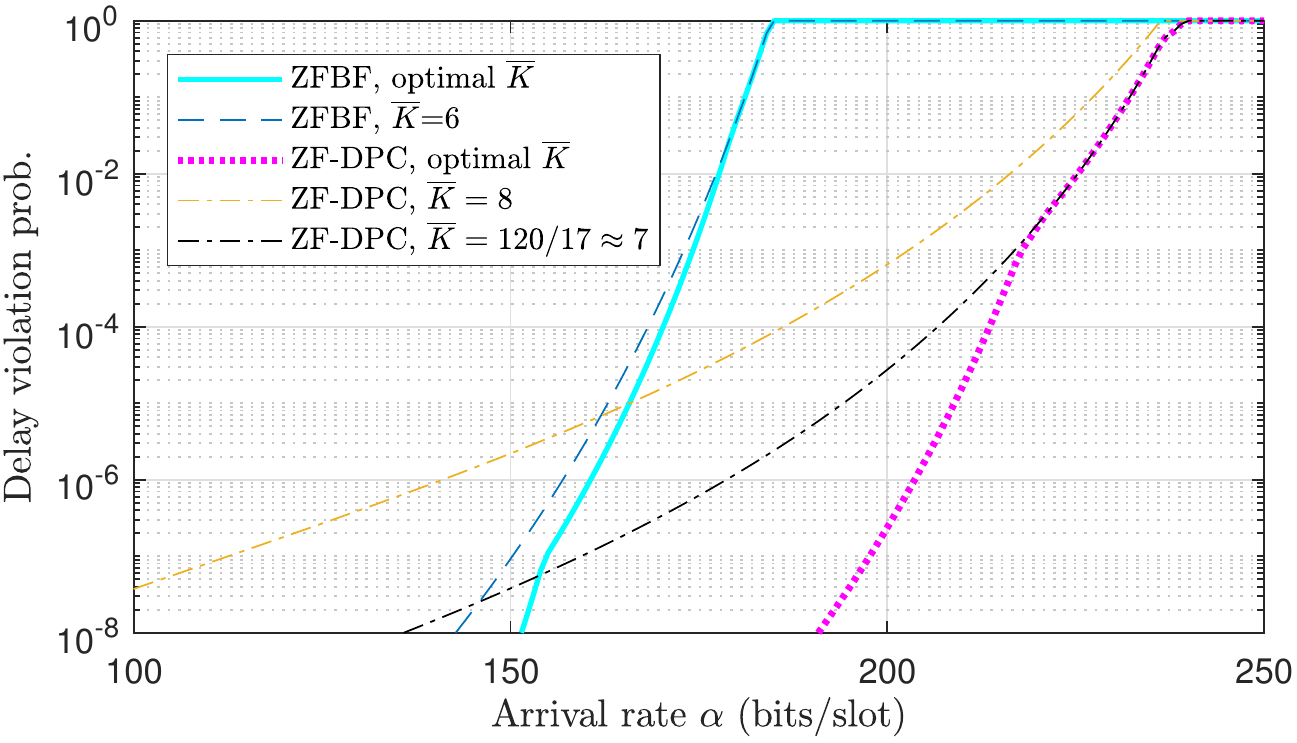}
	}
	\caption{$\Nt=8$ antennas, $\Ktot=120$ users, $\nd=1000$ symbols, $P_\Sigma=15~\mathrm{dB}$, $w=60$. (a): Optimal choice of $\Kavg$ such that the delay violation probability is minimized. (b) Delay violation probability, with optimal $\Kavg$ for each point, along with suboptimal fixed values of $\Kavg$.}
	\label{fig:results_m8_kavg}
\end{figure}

In Fig.~\ref{fig:results_m8_kavg}, we further investigate the optimal value of $\Kavg$ and how the optimal choice of $\Kavg$ influences the delay performance.
Fig.~\ref{subfig:results_M8_opt_Kavg} shows the optimal values for $\Kavg$ for ZFBF (blue, solid lines) and ZF-DPC (red, dotted lines). 
We observe that the optimal value of $\Kavg$ decreases when the arrival rate $\alpha$ is reduced. 
In Fig.~\ref{subfig:results_M8_min_pv}, we investigate how the optimal choice of $\Kavg$ affects the delay performance of the considered systems. 
In case of ZFBF, we find that choosing the suboptimal value $\Kavg=6$ deteriorates the performance only slightly. 
For ZF-DPC, the selected value of $\Kavg$ seems to have a larger impact. We observe that choosing the value $\Kavg=8$, i.e., the value that maximizes the expected service rate $\expected{\Sbit}$ of the system, would lead to a massive increase in the delay violation probability.

\section{Conclusions}
\label{sec:conclusions}
In this work, we have presented an analytical framework to study the delay performance of the multiuser MISO downlink. We found that the optimal number of scheduled users depends on the delay requirements of the system. There are many interesting possible extensions of this work. First of all, we considered equal power allocation, whereas the transmitter could also optimize the transmission power. Another line of research would be to investigate the system performance for a huge number of transmit antennas (massive MIMO). Finally, when the maximum tolerable delay becomes very short, the length of each time slot should also be chosen very small. For very short time slots, the channel estimates may become inaccurate, and the impact of channel coding at finite blocklength must be considered in order to gain more realistic insights into the system performance.

\bibliographystyle{IEEEtran}
\IfFileExists{../../PHD2.bib}
{\bibliography{../../PHD2}}
{
	\IfFileExists{../PHD2.bib}
	{\bibliography{../PHD2}}
	{\bibliography{multiuser_mimo}}
}

\end{document}